\documentclass[11pt,letterpaper]{article}
\usepackage[left=0.88in,right=0.88in,top=0.80in,bottom=0.74in,footskip=24pt]{geometry}
\usepackage[T1]{fontenc}
\usepackage{amsmath,amssymb,amsthm}
\usepackage{txfonts}
\usepackage{booktabs,array,graphicx,microtype,titlesec}
\usepackage[font=small,labelfont=bf,labelsep=period]{caption}
\usepackage[hidelinks,breaklinks=true]{hyperref}
\hypersetup{pdftitle={Admissible Perturbations in Piecewise-Smooth Computation},pdfauthor={Qinyou Wang},pdfsubject={Finite-precision routing certificates and protected parameter updates},pdfdisplaydoctitle=true}
\numberwithin{equation}{section}
\titleformat{\section}{\normalfont\fontsize{12}{14}\selectfont\bfseries}{\thesection}{0.75em}{}
\titleformat{\subsection}{\normalfont\normalsize\bfseries}{\thesubsection}{0.75em}{}
\titleformat{\subsubsection}{\normalfont\normalsize\itshape}{\thesubsubsection}{0.75em}{}
\titlespacing*{\section}{0pt}{2.0ex plus .4ex minus .2ex}{0.8ex plus .2ex}
\titlespacing*{\subsection}{0pt}{1.6ex plus .3ex minus .2ex}{0.6ex plus .2ex}
\titlespacing*{\subsubsection}{0pt}{1.3ex plus .3ex minus .2ex}{0.5ex plus .2ex}
\titlespacing*{\paragraph}{0pt}{1.4ex plus .2ex}{0.75em}
\makeatletter
\renewcommand{\@maketitle}{%
  \begin{center}
    {\fontsize{17}{20}\selectfont\bfseries\@title\par}
    \vskip 12pt
    {\normalsize\@author\par}
  \end{center}\vskip 9pt}
\makeatother

\allowdisplaybreaks[1]
\newtheorem{theorem}{Theorem}[section]
\newtheorem{proposition}[theorem]{Proposition}
\title{Admissible Perturbations in Piecewise-Smooth Computation}
\author{Qinyou Wang}
\date{}
\begin{document}
\maketitle
\begin{abstract}
A parameter update can change a discrete routing decision and invalidate an output bound derived on the original branch. For a finite acyclic graph with comparison-based gates, we give a sufficient condition for a fixed-branch bound to survive native finite-precision routing. We characterize the smallest set of gates that restores a parent trace, when a new write direction strictly increases first-order gain, and which writes preserve protected hidden states after the final router. We test these results on a 30.5-billion-parameter mixture of experts using a function-completion workload. On a protected prefix, a reciprocal expert exchange reproduced and removed a KL violation of $3.24\times10^{-4}$ while the predicted token stayed the same. On another prefix, one exchange accounted for twelve downstream changes in expert membership. At fixed parent, batch and radius, two margin constraints from the failing prefixes yielded an update that lowered sensitive-token training loss by $4.52\times10^{-4}$ and passed all 10,000 protected positions. At a later matched step, a sixth output axis raised the maximum first-order gain by 69.8\% under the same 26 gate constraints. After the final router, ten consecutive updates passed the 10,000-position guard through batch~50 as mean loss on 1,099 fixed sensitive training positions fell from 1.062938 to 1.059590. After 40 training batches, the open-loop checkpoint exceeded the KL cap at 6,734 of 10,000 protected positions and lost 38 of 155 previously passing tests, while the guarded checkpoint passed the full scan and retained all 155.
\end{abstract}
\section{Introduction}
Consider a computation graph whose maps are smooth once its discrete gate decisions are fixed. Near a tie, a change of a few units in the last place of a gate logit can reverse a decision. The selected branch then supplies different inputs to later gates, so a small parameter update can trigger a cascade of switches and a finite jump in output divergence. Enforcing affine lower bounds on gate margins poses a second problem. Their half-spaces can exclude the entire update ball or leave feasible writes with no positive first-order task gain.

These two obstructions lead to distinct geometric questions. Under what conditions does a readout bound derived on a fixed branch carry over to native finite-precision routing? At a fixed radius and with the same margin constraints, which new writable directions strictly increase the attainable first-order gain? A direction orthogonal to the task gradient can still increase that gain by supplying displacement along a constraint normal.

\paragraph{Relation to prior work.}
Sparse MoEs use a discrete decision to route tokens~\cite{shazeer}; differentiable selection and training-time changes to $k$ address the behavior of this boundary~\cite{dselect,chai}. Router margins have been used to preserve expert ordering during quantization~\cite{park}, and route interventions measure damage mediated by quantized routing inputs~\cite{guquant}. Guard-stability analysis bounds when rounded predicates preserve real-arithmetic control flow~\cite{titolo}. Low-rank parameterizations restrict the cost and dimensionality of adaptation~\cite{lora}, while classifier margin bounds certify a decision against bounded changes~\cite{hein}. Our starting point is a smooth-branch bound on protected state evolution~\cite{wang5}. We study its transfer through a causal sequence of finite-precision routing decisions and use native interventions to find the comparisons that matter to a particular failed write.

We test whether successive nonzero writes can lower sensitive-token training loss while preserving the leading token and keeping forward KL from the original model within $2\times10^{-5}$ at every protected position. A zero write at the accepted parent leaves these readouts fixed and gives zero task gain. At any prefix with a unique leading token, the greedy decision persists while the forward KL stays below that prefix's distance to a tie (Eq.~\eqref{eq:greedy-tie}).

Section~\ref{sec:chamber} develops the comparison-graph and subspace results. Section~\ref{sec:benchmark} specializes them to sparse expert writes. Sections~\ref{sec:native-results}--\ref{sec:release} test route restoration, gains from expansion, and the stability of discrete readouts.

\section{Chamber lifting}
\label{sec:chamber}
Theorem~\ref{thm:branch-transfer} lifts a readout bound on a fixed branch to native finite-precision execution. Enlarging the writable space is a second operation. Proposition~\ref{prop:writable-gain} characterizes strict gain under this enlargement, and Section~\ref{sec:terminal-support} constructs its norm-preserving embedding for sparse writes.
\subsection{From a fixed branch to native execution}
\label{sec:route-membership}

Let $W_{\rm par}$ be the current parent, $W_0$ the immutable reference for protected behavior, and $x\in B_\rho=\{x\in\mathbb R^d:\|x\|\le\rho\}$ a physical parameter increment about $W_{\rm par}$. The norm is Euclidean on coefficient vectors (Frobenius on their matrix representations). Fix finitely many inputs and their protected readouts. A finite deterministic acyclic computation graph routes at each gate $v$ by finitely many score comparisons and a specified tie rule. Its branch is the trace of discrete labels on the executed graph. Forcing the parent's labels $S_v$ defines the parent branch throughout $B_\rho$, including at parent ties. The continuous states and scores on this branch still depend on $x$. The trace includes decisions that enable subsequent subgraphs. At a top-$k$ gate, $S_v$ records the selected indices and any ordering that affects the computation.

For each comparison $j$ needed to determine the parent label at $v$, let $h_{vj}$ be its real score gap oriented toward that decision, and $\widehat h_{vj}$ its native value conditional on parent labels upstream. Assume $|\widehat h_{vj}-h_{vj}|\le e_{vj}$ in dispatcher-score units and $\|\nabla^2 h_{vj}\|_{\rm op}\le L_{vj}$ on the forced branch throughout the ball. For unordered top-$k$ sets, the comparisons separate selected from excluded indices and the bounds cover the possible native reduction orders.

\begin{theorem}[Conditional chamber lifting]
\label{thm:branch-transfer}
Fix $x\in B_\rho$. Suppose every defining comparison on the protected causal graph either has an independent certificate of the parent's decision (including the native tie decision at equality), or satisfies
\begin{equation}
 h_{vj}(0)+\nabla h_{vj}(0)^\top x
      >\tfrac12L_{vj}\|x\|^2+e_{vj}.
 \label{eq:gate-cert}
\end{equation}
Then native execution at $x$ reproduces the parent trace. If the fixed-branch divergence from $W_0$ at anchor $a$ is bounded by $B_a(x)$ and its native evaluation on the branch incurs absolute error at most $e_a^{\rm out}$, then $D_a^{\rm native}(x)\le B_a(x)+e_a^{\rm out}$.
\end{theorem}
\begin{proof}
Visit the gates in topological order. Given the parent labels upstream, Taylor's lower bound and Eq.~\eqref{eq:gate-cert} give a positive native gap at $v$. The independent certificates settle every remaining pair, which establishes $S_v$ and closes the induction. The native endpoint therefore uses the certified branch, where the readout bound applies.
\end{proof}

The transfer covers any scalar discrepancy bounded on the branch. At a fixed observation, for example, two Gaussian policies with common positive definite covariance $\Sigma$ have KL $\tfrac12\|\Sigma^{-1/2}(\mu_x-\mu_0)\|^2$.

\paragraph{A raw-logit floor and native dispatch.}
Let $\delta=r_i-r_j$ be the real router-logit gap and $p=\operatorname{softmax}(r)$.
The score gap in dispatcher units is $p_i-p_j=p_j(\exp(\delta)-1)$.
If $|\widehat p_k-p_k|\le\epsilon_k$ for $k=i,j$, a gap exceeding
$\epsilon_i+\epsilon_j$ certifies strict native order.
These score errors include upstream arithmetic on the branch.
At equality of the rounded scores, certification uses the dispatcher's actual tie decision.

\paragraph{Restoring a routing trace.}
Fix an input, a candidate write $x$, and the parent execution graph for its readout. A clamp supplies the parent's discrete choice. The endpoint computes the continuous values on that branch, including any routing weights. Assume deterministic dispatch, including its tie rule, and that forcing a gate's own native label leaves its calculation unchanged. At candidate weights, replay the graph with every gate clamped to its parent label. Let $F(x)$ contain the gates whose native choice, recorded immediately before the override, differs from that label.

\begin{proposition}[Minimal restoration set at a fixed write]
\label{prop:restoration}
Clamping a set $C$ to its parent labels and running the remaining gates natively reproduces the entire parent routing signature if and only if $F(x)\subseteq C$.
\end{proposition}
\begin{proof}
With parent labels upstream, each gate receives exactly the inputs it received in the all-clamped candidate replay. If $F(x)\subseteq C$, every free gate therefore selects its parent label by topological induction. Conversely, if the whole signature is restored, each free gate saw these same inputs and chose its parent label, so every member of $F(x)$ was clamped.
\end{proof}
Thus $F(x)$ is the unique inclusion-minimal restoration set. The earliest naturally changed gates belong to it. A downstream gate can also belong if its choice changes even with its ancestors restored.

\paragraph{Reciprocal intervention at one readout.}
For parameter states $b\in\{0,1\}$ (parent, candidate), let $D_{bs}$ be the protected readout divergence from $W_0$ with one gate fixed to label $s\in\{0,1\}$ (parent, candidate) and subsequent gates evaluated natively. The diagonal entries reproduce the native parent and candidate by consistency. Given a passing parent and a failing candidate, the off-diagonal entries test whether the exchange induces the focal violation at the parent and removes it at the candidate. Both tests pass exactly when
\[
  \max_b D_{b0}\le \bar D < \min_b D_{b1}.
\]

\paragraph{Cumulative output protection.}
\label{sec:fixed-branch-kl}
Let $z_{a,0}$ be the stored reference logits, $z_{a,\rm par}$ the real parent-branch logits, $J_a$ their Jacobian at the parent, and $\Pi_\circ=I-\mathbf1\mathbf1^\top/V$ the centering projector for a categorical readout with $V$ outcomes. After route certification, let $E_a(x)$ bound the centered Euclidean discrepancy between the native endpoint logits and $z_{a,\rm par}+J_ax$. This includes the smooth remainder ($L_a\|x\|^2/2$ for an $L_a$-Lipschitz centered Jacobian along $[0,x]$), real-to-native error and any rounding of the affine prediction. For temperature $T>0$, the exact softmax KL $K_a$ from the stored reference logits to the native endpoint logits satisfies
\begin{equation}
 K_a(x)\le\frac{
 \big(\|\Pi_\circ(z_{a,\rm par}-z_{a,0}+J_ax)\|+E_a(x)\big)^2}{4T^2}.
 \label{eq:lifted-logit}
\end{equation}
Indeed, softmax KL is the Bregman divergence of log-sum-exp in scaled logits. Its Hessian in those coordinates is a categorical covariance. A unit vector $u$ has variance at most $(\max_i u_i-\min_i u_i)^2/4\le1/2$. Taylor's bound and invariance to a constant logit shift give Eq.~\eqref{eq:lifted-logit}~\cite[Cor.~11.2]{wang5}. The parent offset accounts for earlier accepted writes. Rounded computation of the KL scalar contributes its own evaluation error. 
\paragraph{Distance to a greedy tie.}
At a fixed input, let label $t$ lead runner-up $s$ by $\gamma=z_t-z_s>0$.
Any rival $j$ reaching $t$ after a logit change satisfies
$(e_j-e_t)^\top\Pi_\circ\Delta z\ge\gamma$, hence
$\|\Pi_\circ\Delta z\|\ge\gamma/\sqrt2$.
The forward-KL threshold uses the closest competing mode of a categorical distribution~\cite[App.~D]{pacmode}.
Write $a=p_t$, $b=p_s$ for the reference softmax probabilities. Then
\begin{equation}\label{eq:greedy-tie}
 \min_{q:\,\max_{j\ne t}q_j\ge q_t}D_{\rm KL}(p\|q)
 =a\log\frac{2a}{a+b}+b\log\frac{2b}{a+b}.
\end{equation}
For a rival of mass $u$, group the pair in the KL chain rule.
The aggregate term is minimized by retaining its mass $a+u$ and all other masses.
The conditional binary divergence is minimized at a tie. The resulting cost has derivative
$\log(2u/(a+u))\le0$ in $u$, so the runner-up attains the minimum.

\begin{figure}[tbp]
\centering
\includegraphics[width=\linewidth]{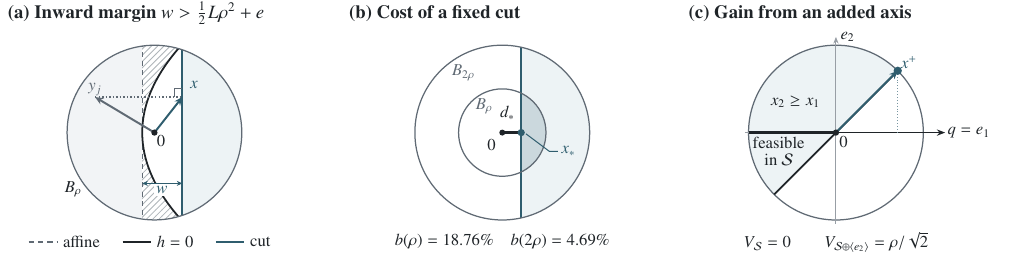}
\caption{\textbf{Route margins and writable directions.}
(a) For $h(u,v)=u+0.24-0.26v^2$, hatching marks the region the affine gap classifies as positive while $h<0$.
The cut $u\ge0.55$ lies $w=0.79$ from the affine boundary. With unit gradient at the origin, $L=0.52$, $\rho=1.72$ and $e=0.02$, this exceeds $L\rho^2/2+e$. The dotted segment projects $y_j$ onto the cut.
(b) The shaded caps are feasible. The percentages give the squared-radius cost $d_*^2/r^2$ for the two gate rows at levels 16 and 15.
(c) For $q=e_1$, the best write on the old axis is $0$. Adding $e_2$ permits $x^+=(\rho/\sqrt2,\rho/\sqrt2)$.}
\label{fig:geometry}
\end{figure}

\subsection{The cost of gate constraints and the value of a new axis}
\label{sec:affine-budget}
An affine lower bound on a gate margin restricts the write to a half-space. Figure~\ref{fig:geometry} illustrates how the radius changes the physical cost of a fixed floor.

\begin{proposition}[Budget for fixed gate rows]
\label{prop:gate-budget}
Let the fixed rows $A_Rx\ge c_R$ be feasible and set $d_*=\min\{\|x\|:A_Rx\ge c_R\}$. The ball meets these rows if and only if $\rho\ge d_*$. For $\rho>0$ in this range, the least squared-radius fraction needed by these rows is
\begin{equation}
 b(\rho)=d_*^2/\rho^2,\qquad b(2\rho)=b(\rho)/4.
 \label{eq:gate-budget}
\end{equation}
\end{proposition}
\begin{proof}
The minimum-norm point of the closed feasible polyhedron belongs to every such ball. Every feasible write has norm at least $d_*$. Dividing the minimum squared norm by the squared radius gives the fraction. In particular, any row $a_i^\top x\ge c_i>0$ imposes $d_*\ge c_i/\|a_i\|$ by Cauchy--Schwarz.
\end{proof}

Fix the parent, training batch and physical inner product. Let $q$ be the negative task-loss gradient, and collect the affine gate and output rows as $Ax\ge c$. The dual problem below has one multiplier per retained row. Building the rows and checking the endpoint require evaluations on the protected graph. A top-$k$ gate on $N$ scores has $k(N-k)$ selected--excluded pairs before any comparisons are grouped or certified together.

\begin{proposition}[Gain from enlarging the write space]
\label{prop:writable-gain}
Let $\mathcal S$ be a writable subspace with orthogonal projector $P_{\mathcal S}$. Suppose some $x\in\mathcal S$ satisfies $Ax>c$ and $\|x\|<\rho$. Then
\begin{equation}
 \begin{aligned}
 V_{\mathcal S}(\rho)&=\max\{q^\top x:x\in\mathcal S,\ \|x\|\le\rho,\ Ax\ge c\}\\
 &=\min_{\lambda\ge0}\{\rho\|P_{\mathcal S}(q+A^\top\lambda)\|-c^\top\lambda\}.
 \end{aligned}
 \label{eq:writable-gain}
\end{equation}
Write $\Lambda_{\mathcal S}^*$ for the set of minimizing multipliers. With all old directions writable and $\mathcal E\perp\mathcal S$, enlargement strictly raises $V$ if and only if $P_{\mathcal E}(q+A^\top\lambda)\ne0$ for every $\lambda\in\Lambda_{\mathcal S}^*$.
\end{proposition}
\begin{proof}
Maximizing $q^\top x+\lambda^\top(Ax-c)$ over the subspace ball gives the dual objective. Strict feasibility ensures equality and dual attainment in both spaces~\cite{boyd}. Orthogonality adds $\|P_{\mathcal E}(q+A^\top\lambda)\|^2$ to the squared norm. An old minimizer with zero added component certifies equality. Conversely, if the values agree, any new minimizer attains the old minimum and its added component vanishes.
\end{proof}
An added direction can carry a protection constraint even when its task derivative is zero. For $q=e_1$, $\mathcal S=\operatorname{span}(e_1)$ and $x_2-x_1\ge0$, the old maximum is zero. Adding $e_2$ permits $x=(\rho/\sqrt2,\rho/\sqrt2)$, yielding gain $\rho/\sqrt2$ despite the zero task derivative along $e_2$.

\section{Protected writes in a sparse language model}
\label{sec:benchmark}

\subsection{A sparse expert write}
\label{sec:parameterization}
At layer $\ell$ of an $L$-layer network, let $S_\ell(h)$ select $k$ of $N$ experts and let $E$ be the set of $m$ writable experts. For a shared orthonormal output basis $U\in\mathbb R^{d_{\rm out}\times r}$ and input coefficients $\Theta_e\in\mathbb R^{r\times d_{\rm in}}$, add to the frozen mixture
\[
    \Delta f_\ell(h)=\sum_{e\in S_\ell(h)\cap E} w_e(h)\,U\Theta_e h ,
\]
using the native normalized weights $w_e(h)$ (Fig.~\ref{fig:controller}a). The controller proposes a write in these coordinates, checks the stored endpoint on protected inputs and commits an accepted write. At a terminal router, the fraction of protected positions outside the support is $|A_g|^{-1}\sum_{a\in A_g}\mathbf 1\{S_\ell(h_a)\cap E=\varnothing\}$ on a panel $A_g$. It depends on which experts those inputs select.

\subsection{Neutral expansion and writes after the last router}
\label{sec:terminal-support}
Appending orthonormal columns perpendicular to $U$ embeds every old
coefficient increment as $\Delta\Theta_e\mapsto(\Delta\Theta_e^\top,0)^\top$.
This preserves the physical Frobenius norm and each expert's real output,
so the original feasible writes embed isometrically when the same constraints
are imposed on the physical write and its readouts. A native implementation
can preserve the old computation bit for bit by bypassing a zero new branch
and retaining the original kernels and reduction order. We check this equality on the protected panel at installation.

\begin{proposition}[Terminal sparse support kernel]
\label{prop:terminal-kernel}
Fix the input prefixes, all upstream parameters and an $r$-column basis
$U$ of full column rank. Place the write after the last router and attention
operation, with only positionwise operations thereafter. Write $d=d_{\rm in}$. At readout $a$ let
$h_a\in\mathbb R^d$ be the expert input and $w_{ae}$ the routing weight
of writable expert $e$, set to zero when unselected. Stack
$\phi_a=(w_{ae}h_a)_{e=1}^m\in\mathbb R^{md}$, collect the protected columns in
$\Phi=[\phi_a]_a$, and concatenate coefficient increments horizontally
in $\Delta B\in\mathbb R^{r\times md}$. The real writes preserving all
protected post-mixture hidden states are precisely
\[
 \{\Delta B:\Delta B\Phi=0\},\qquad
 \dim=r(md-\operatorname{rank}\Phi).
\]
\end{proposition}
\begin{proof}
All router inputs precede the write, so the features $\phi_a$ are fixed.
The hidden increment is $U\Delta B\phi_a$. Injectivity of $U$ gives the
stated kernel, and rank--nullity applies independently to each row of
$\Delta B$. Positionwise downstream execution carries each preserved hidden
state to the same logits.
\end{proof}
There are no downstream gate rows, so their minimum-norm cost is zero.
At an anchor selecting no writable expert, the native implementation copies
the original mixture directly. With the same deterministic readout path
its logits, and hence its divergence from $W_0$, are unchanged for every
finite write for which execution is defined. For cancellations between active experts, native equality requires a separate floating-point check.

\paragraph{Relation to MoEEdit.}
The preservation term in MoEEdit has the form $\|\Delta W\Psi\|_{\rm F}^2$, where $\Psi$ stacks routing-weighted expert activations~\cite[Eq.~(5)]{moeedit}.
MoEEdit updates expert output projections using per-expert near-null eigenspaces.
In our terminal construction, the shared-input features $\Phi$ remain fixed under the write, so $\Delta B\Phi=0$ characterizes all finite coefficient updates that preserve the protected hidden states in real arithmetic, including cancellations between active experts.

\subsection{Model, workload and controller}
\paragraph{Model and writable experts.}
We instantiate this construction on Qwen3-30B-A3B-Base~\cite{qwen}, a 30.5-billion-parameter model with 48 layers, $N=128$, $k=8$ and $d_{\rm in}=d_{\rm out}=2{,}048$. At layer 23 (zero-based), we choose the eight experts with the greatest routing load on protected prefixes, weighting each token by the number of protected readouts it can influence. This selection stresses the guard at sites frequently used by protected prefixes, with routers 24--47 downstream. On one RTX PRO 6000 Blackwell Server Edition, frozen weights reside in BF16 and are converted to FP32 on demand. Forward and derivative computations disable TF32 and autocast. The dispatcher compares FP32 softmax scores.

\paragraph{Training objective and discrete readouts.}
SDK440 supplies an earlier caller and a new interface signature for each completion task. We train on 160 tasks from 20 families and greedily decode 40 held-out tasks from five other families. Following execution-based evaluation~\cite{chen,evalplus}, our deterministic Python harness checks return values and specified call sequences on 312 fixtures. Of 20 same-domain legacy clients, ten passed every fixture at $W_0$. We froze their 59 fixtures before checkpoint replay. Together with the 96 held-out fixtures passed by $W_0$, they give 155 baseline passes.

\paragraph{A fixed reference for each update.}
Each of 1,000 fixed documents supplies ten protected positions
with zero-based indices 6, 19, 31, 44, 57, 69, 82, 95, 107 and 120. At every position $a$ in this panel $A_g$,
a proposed checkpoint must preserve the reference top token and satisfy
$D_{\rm KL}(p_{W_0}(\cdot\mid a)\Vert p_{\rm cand}(\cdot\mid a))
\le 2\times10^{-5}$ over the full vocabulary. The fixed reference measures cumulative drift. We replay trajectories at designated checkpoints after step selection.

\begin{figure}[tbp]
\centering
\includegraphics[width=.99\linewidth]{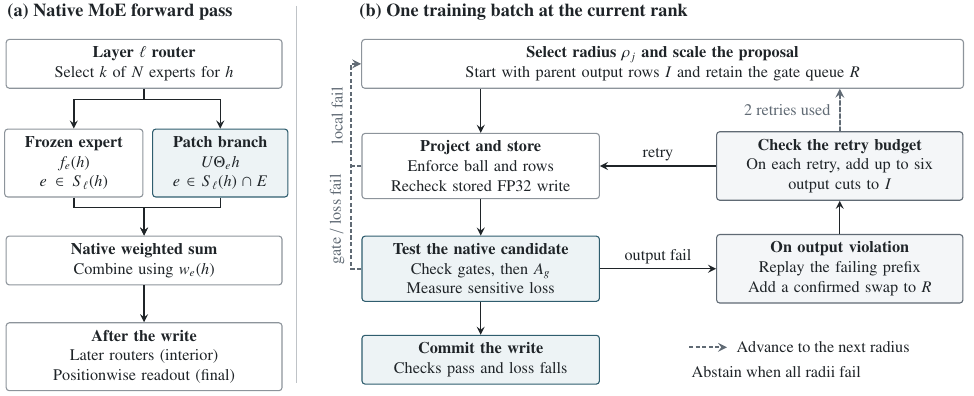}
\caption{\textbf{Patch placement and radius search.}
(a) Only experts in $S_\ell(h)\cap E$ contribute a patch. After the final mixture the readout acts at each position separately.
(b) First-admissible search at rank five (batches 7--20). Dashed arrows advance the radius after a failed local or tracked-gate check, a nonpositive gain, or two retries. An output violation can add a gate to $R$ through reciprocal replay. Each retry adds output cuts to $I$.}
\label{fig:controller}
\end{figure}

\paragraph{Four axes and a neutral fifth.}
The 160 training tasks contain 25,622 target tokens, including 1,099
execution-sensitive positions. The initial objective weights each sensitive
position fourfold and normalizes the token weights to mean one. At $W_0$,
we differentiate the family losses with respect to full expert write
matrices and right-project the gradients off the protected input-Gram
eigenvectors with eigenvalues above $10^{-6}$. The four shared output axes
comprise one normalized Gaussian vector and the three leading directions
of the family-averaged output gradient second moment in its orthogonal
complement. A separate sensitive-position moment $S_{\rm sens}$ gives
equal weight to families and, within each family, to observed decision
stages. We fit three directions in the same Gaussian axis's orthogonal
complement, project them off the four-axis span and
orthonormalize the residuals into $V$. The fifth axis is $Vq$, where $q$
is a leading unit eigenvector of $V^\top S_{\rm sens}V$. Its coefficients
start at zero. At rank five, the eight experts carry 81,920 writable FP32
coefficients (320 KiB), plus 40 KiB for the shared basis. The deployed patch uses the native top-$k$ dispatcher.

\paragraph{Candidate construction and native acceptance.}
\label{sec:method}
The static four-axis comparator commits 40 AdamW updates in batches of
eight tasks, at learning rate $0.003$ and zero weight decay. In the five-axis continuation of Section~\ref{sec:native-results}, the controller projects a scaled Adam proposal $y_j$ at physical radius
$\rho_j$, with $\|x\|_2^2=\sum_e\|U\Delta\Theta_e\|_{\rm F}^2$.
For retained output rows $i\in I$ and tracked gate pairs $v\in R$, it solves
\begin{equation}
\begin{aligned}
  \min_{\|x\|_2\le\rho_j}\quad &\tfrac12\|x-y_j\|_2^2 \\
  \text{subject to}\quad
  &f_i(x_i)+\nabla f_i(x_i)^\top(x-x_i)\le \ell_i
       &&(i\in I),\\
  &h_v(0)+\nabla h_v(0)^\top x\ge \tau_v s_v &&(v\in R).
\end{aligned}
\label{eq:local-gate-solve}
\end{equation}
The vector $x_i$ is the write at which row $i$ is linearized, equal to zero for a parent row and the violating stored write for a retry cut. Its limit $\ell_i$ is $2\times10^{-5}$ for cumulative KL or zero for a rival's logit minus the reference winner's. A gate row uses the retained expert's raw logit minus its rival's.
Here $s_v$ is one FP32 unit in the last place (ULP) at the magnitude of
the eighth-ranked parent logit. The two batch-six rows use $\tau_v=6$. These empirical floors guide proposals. The a priori test in Theorem~\ref{thm:branch-transfer} additionally requires bounds on curvature and native score error.

After FP64 projection and FP32 storage, the controller (Fig.~\ref{fig:controller}b) rechecks the realized
radius and rows, then checks tracked expert membership and positive raw
gaps under native dispatch. It commits a write only after all 10,000
outputs pass and same-batch sensitive-token negative log-likelihood (NLL)
decreases. This scan tests each protected readout, including those whose untracked upstream assignments changed. For the first failing anchor, the controller orders prefix membership changes by layer, then token, skipping tracked pairs. At the first remaining change it tests a single outgoing/incoming expert pair in both directions. A compound exchange yields no row. A new gate is retained when the swap induces the focal violation at the parent and removes it at the candidate. Up to
two same-radius retries add offending output rows. Batches 7--20 commit
the first admissible write. Batches 21--25 also test the next three larger
dyadic radii and choose the admissible candidate with the largest measured
loss decrease.
\section{Identifying a routing failure and recovering the update}
\label{sec:native-results}

\subsection{Two reciprocal interventions}
On batch~6 the five-axis controller tested 47 distinct coefficient tensors and retained its batch-5 parent. At each of five successive halving levels, both writes proposed on the same ray before coefficient rounding produced a KL divergence of approximately $3.24\times10^{-4}$ at token~69 of document~753. The parent's divergence at the same position was $2.215\times10^{-12}$. Tracing the 70-token prefix located a change at the readout token in layer~33. Experts~50 and~93 had been separated in the parent by two units in the last place (ULPs) of the local FP32 router logit.

We tested the effect of that assignment in both directions. For each endpoint we replaced the expert in the last selected slot and let all subsequent routers execute natively. Replaying each endpoint's own assignment reproduced its logits bit for bit. Selecting expert~93 in the parent reproduced the violation. Restoring expert~50 in the rejected candidate brought the focal position below the guard limit (Table~\ref{tab:gate-interventions}). The two columns fall on opposite sides of the KL cap in both rows. All four executions predicted token~15. A check on the predicted token alone would have accepted both assignments.

\begin{table}[htbp]
\centering\small
\caption{One expert assignment switches the KL at document~753, token~69. Entries give $D_{\mathrm{KL}}(p_{W_0}\Vert p)$ at the original level-16 trial. Bold marks native dispatch.}
\label{tab:gate-interventions}
\begin{tabular}{lrr}
\toprule
Coefficients & Select expert 50 & Select expert 93 \\
\midrule
Batch-5 parent & $\boldsymbol{2.215\times10^{-12}}$ & $3.240\times10^{-4}$ \\
Rejected candidate & $9.103\times10^{-12}$ & $\boldsymbol{3.240\times10^{-4}}$ \\
\bottomrule
\end{tabular}
\end{table}

\begin{figure}[htbp]
\centering
\includegraphics[width=.92\linewidth]{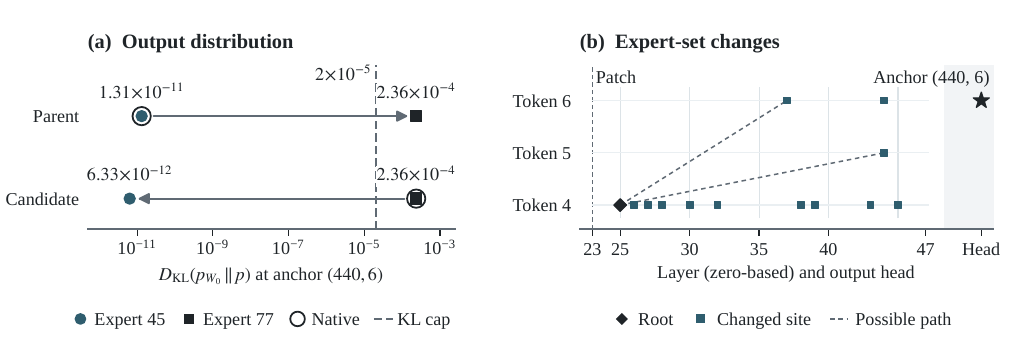}
\caption{Reciprocal intervention on document~440. (a) Rings mark the native assignments. Arrows show the focal KL after exchanging experts~45 and~77. (b) The update is at layer~23, the exchanged assignment is at layer~25, token~4 (diamond), and the readout is at token~6 after layer~47 (star). Squares mark the twelve downstream membership changes. Dashed guides show possible cross-token paths through causal attention.}
\label{fig:root-gate}
\end{figure}

We added a constraint on the 50--93 margin and repeated the solve at the same radius. Expert~50 remained selected and focal KL fell to $1.579\times10^{-11}$. The scan then failed at token~6 of document~440. Here the initiating gate was two tokens earlier. The batch-5 parent selected expert~45 in an exact raw-logit tie at layer~25, token~4, while the candidate favored expert~77 by two ULPs. The selected expert sets differed at 13 of the prefix's 336 router sites. Restoring expert~45 in the candidate recovered the parent's memberships at every site. The reverse exchange reproduced all the candidate's memberships in the parent (Fig.~\ref{fig:root-gate}). This one clamp restored the complete measured membership trace, including twelve subsequent switches. The empty set left the thirteen observed differences (Proposition~\ref{prop:restoration}). All four runs predicted token~304 at the readout.

\subsection{Repair at a fixed physical radius}
\label{sec:radius-sweep}
Of fifteen larger scheduled radii tested with both gate rows, level~9 alone passed the full guard and lowered training loss. At that radius we compared solves with zero, one and two gate rows, holding the batch-5 parent, the eight training tasks of batch~6 and the physical radius fixed at $8.69033\times10^{-4}$ (level~9). Table~\ref{tab:level-nine} shows why both rows mattered. The first row brought the focal KL at document~753 back below the cap. The second admitted a write that passed the complete guard, with all 10,000 top tokens preserved. Its measured decrease in same-batch sensitive-token NLL was $4.52377\times10^{-4}$.

\begin{table}[htbp]
\centering\small
\caption{Matched level-9 writes. Each failing scan stopped after the first document with a violation. The maximum in each row is taken over the indicated number of checked anchors.}
\label{tab:level-nine}
\setlength{\tabcolsep}{5pt}
\begin{tabular}{lrrrl}
\toprule
Gate rows & KL at $(753,69)$ & Anchors checked & Max. KL checked & Outcome \\
\midrule
None & $3.2392\times10^{-4}$ & 10 & $3.2392\times10^{-4}$ & Failed at $(753,69)$ \\
753 & $4.0211\times10^{-12}$ & 60 & $2.3620\times10^{-4}$ & Failed at $(440,6)$ \\
753 and 440 & $9.1344\times10^{-12}$ & 10,000 & $9.1204\times10^{-6}$ & Accepted \\
\bottomrule
\end{tabular}
\end{table}

For these two fixed gate rows, the minimum-norm correction used 18.76\% of the squared radius at level~16 and 4.69\% at level~15, where the radius doubled (Fig.~\ref{fig:geometry}b). At the accepted level-9 endpoint the two affine raw-logit predictions were both approximately $+6$ ULPs. Native evaluation gave $+1$ ULP in document~753 and $+5$ in document~440. The difference matters even within this batch. At the smaller level-11 radius, a fresh projection also predicted $+6$ ULPs at 753, but the native gap was $-1$ ULP. Expert~93 entered and the violation recurred. A fresh projection changes the direction as well as the length of a write, so native acceptance can vary nonmonotonically across radii.

\subsection{Continuation across nineteen batches}
\label{sec:online-continuation}
The recovered checkpoint admitted a write that lowered same-batch sensitive-token NLL on each of the remaining 14 batches of the first epoch. Across 43 proposals, 16 failed a local test, two failed native tracked-gate checks, and eleven failed the native output scan. Each of those eleven output failures yielded a reciprocal expert exchange that reproduced the focal violation in the parent and restored focal compliance in the candidate. On the first five batches of the second epoch the controller tested three larger radii after finding its first admissible write. It selected a larger radius on three batches. Over all 19 committed endpoints the largest KL relative to $W_0$ was $1.41104\times10^{-5}$, and every scan preserved all 10,000 top tokens.

Batch~25 yielded a separate same-radius test of constraint retention. Adding a gate for document~18 evicted an older row for document~351. Replaying the eight-row proposal from the batch-24 parent stopped after 8,310 anchors with KL $4.16280\times10^{-5}$. We reinstated the evicted row, holding the parent, proposal, radius and 16 output rows fixed. The native expert-81 minus expert-68 gap at the tracked site moved from $-2$ to $+4$ ULPs. The nine-row write passed the full scan, with maximum KL $1.40980\times10^{-5}$, and lowered sensitive-token NLL by $6.82241\times10^{-5}$ on the 50 sensitive training positions.
\section{Write-space expansion and downstream gate constraints}
\label{sec:expansion}

The route intervention changes the feasible rows at fixed output rank. Proposition~\ref{prop:writable-gain} predicts a different effect when a new axis expands the writable space at fixed rows. We tested that distinction at the layer-23 batch-24 parent, with the eight batch-25 training tasks and a physical radius of $4.01683\times10^{-4}$. We took an output direction from the training sensitive-loss gradient remaining after projection off the five existing axes. All five original axes stayed writable. Installing the sixth with zero coefficients preserved all 10,000 protected logit vectors bit for bit. Each additional shared axis supplies $8\times 2{,}048$ expert-specific input coefficients.

\begin{table}[htbp]
\centering\small
\caption{Maximum first-order decrease of the same batch-25 sensitive loss. Within a row the physical radius, objective, parent and protection constraints are identical. Both subspaces face the same 16 output rows. The 26-row configurations retain nine additional gate comparisons. Values are shown where feasible primal and dual bounds agree to the printed precision.}
\label{tab:six}\begin{tabular}{lrrr}\toprule
Gate constraints & Five axes & Six axes & Support ratio\\\midrule
17 & $2.60109\times10^{-4}$ & $4.47277\times10^{-4}$ & $1.720$\\
26, original floors & $2.60099\times10^{-4}$ & $4.41564\times10^{-4}$ & $1.698$\\
26, one floor strengthened & $2.60053\times10^{-4}$ & $4.41527\times10^{-4}$ & $1.698$\\\bottomrule
\end{tabular}\end{table}

Table~\ref{tab:six} gives the support values. For the six-axis trial we materialized the optimizer of Eq.~\eqref{eq:writable-gain}. At the original 17 rows, the six-axis candidate exceeded the KL cap at eleven anchors across nine documents. Forcing the parent assignments on each of those prefixes brought all eleven below $4.35\times10^{-9}$. At each failing anchor a single reciprocal expert swap reproduced the violation in the parent and restored compliance in the candidate. Two documents each had one upstream swap shared by two readouts. Adding the nine distinct gate rows retained $98.72\%$ of the six-axis support. Measuring all eleven violations together allowed the nine constraints to enter a single solve.

The six-axis candidate retained all 26 tracked expert assignments, with one raw-logit tie. Raising that floor from six to nine parent ULPs yielded a native gap of five ULPs. The resulting write passed all 10,000 protected positions (maximum KL $1.77647\times10^{-5}$) and reduced mean NLL on the batch's 50 sensitive positions by $2.57441\times10^{-4}$.

\paragraph{A gate floor outside the search ball.}
Starting from this accepted six-axis checkpoint, the layer-23 run committed writes on batches 26--28 and retained those weights through batch~40. The first batch-30 trial failed a native gate check. Once the floor for document~896 (layer~40, token~107) was raised, its half-space lay $1.04170\times10^{-3}$ from the origin. The three radii were at most $4.01683\times10^{-4}$, so this row excluded all 33 subsequent search balls through batch~40 (Proposition~\ref{prop:gate-budget}).

\paragraph{Writing after the final router.}
We froze the batch-40 checkpoint and placed a new write after the layer-47 router. With no downstream gate rows, their minimum-norm cost is zero (Proposition~\ref{prop:terminal-kernel}). Let $h_t$ denote the normalized input at this layer and $d_t$ the sensitive-loss derivative with respect to its mixture output. We ranked experts by $\sum_t w_{te}\|h_t\|\|d_t\|/(1/128+p_e+p_e^{\rm hot})$ over the 1,099 sensitive training positions. Here $p_e$ and $p_e^{\rm hot}$ are each expert's share of routing weight over all protected readouts and over those with parent KL at least $1.6\times10^{-5}$, respectively. The top eight were 126, 5, 116, 25, 97, 98, 1 and 50. We took the leading eight eigenvectors of $\sum_t\|h_t\|^2(\sum_{e\in E}w_{te}^2)d_td_t^\top$ as the shared output basis.

At all ten readouts of 416 of the 1,000 documents, the last router selected none of these experts. Thus at least 4,160 of the 10,000 columns of $\Phi$ vanish. Proposition~\ref{prop:terminal-kernel} gives $\operatorname{rank}\Phi\le5{,}840$ and at least $8(16{,}384-5{,}840)=84{,}352$ real coefficient directions preserving every protected post-mixture state. Zero-coefficient installation preserved all 10,000 protected logit vectors bit for bit. Batches 41--45 searched along normalized sensitive-loss gradients. Batches 46--50 also tested narrow-margin directions and their projections against high-KL anchor normals. The first accepted write had radius $3.21347\times10^{-3}$, eight times the starting search radius, and reduced mean NLL on the batch's 63 sensitive positions by $3.20275\times10^{-3}$. All ten writes decreased same-batch sensitive-token NLL. On the same 1,099 training positions, mean NLL was 1.062938 at batch~40, 1.060084 at batch~45 and 1.059590 at batch~50. Batch~50 passed all 10,000 protected readouts with maximum KL $1.99737\times10^{-5}$.

\section{Discrete trajectory stability}
\label{sec:release}
After the same 40 training batches, the open-loop checkpoint exceeded the KL cap at 6,734 of 10,000 protected positions, lost 38 previously passing tests (31 held-out and seven legacy), and gained 18 previously failing held-out tests. The guarded layer-23 updates lowered same-batch sensitive-token loss, and the batch-40 checkpoint passed the full scan and preserved all 371 fixture pass/fail outcomes, including all 155 baseline passes.

\paragraph{Greedy trajectories.}
Two deterministic greedy runs share a prefix up to their first differing token. For a unique reference winner, Eq.~\eqref{eq:greedy-tie} gives the least forward KL at that prefix at which a rival can tie it. Sensitive-token NLL averages a separate set of training positions. At the five-axis batch-25 checkpoint all 50 generated sources matched $W_0$ byte for byte. Between the accepted six-axis checkpoint at batch~25 (Section~\ref{sec:expansion}) and batch~40, one of the 40 held-out programs changed text and all fixture outcomes persisted.

\end{document}